\documentclass[prl,twocolumn,showpacs,superscriptaddress,nofootinbib]{revtex4-1}
\usepackage{graphicx}% Include figure files
\usepackage{bm}% bold math
\usepackage{amssymb}
\usepackage{color}
\usepackage{epstopdf}
\usepackage{amsmath}
\usepackage{amstext}
\usepackage{latexsym}
\usepackage[colorlinks=true,citecolor=blue,linkcolor=blue]{hyperref}
\usepackage{mathptmx} %comapct fonts
\def\la{\langle}
\def\ra{\rangle}

\newcommand{\beq}{\begin{equation}}
\newcommand{\eeq}{\end{equation}}
\newcommand{\beqa}{\begin{eqnarray}}
\newcommand{\eeqa}{\end{eqnarray}}

\begin{document}

\title{Tuning heat transport in trapped-ion chains across a structural phase transition}

\author{A. Ruiz}
\affiliation{Departamento de F\'isica, Universidad de La Laguna, La Laguna 38203, Spain}
\affiliation{IUdEA Instituto Universitario de Estudios Avanzados, Universidad de La Laguna, La Laguna 38203, Spain}
\author{D. Alonso}
\affiliation{Departamento de F\'isica, Universidad de La Laguna, La Laguna 38203, Spain}
\affiliation{IUdEA Instituto Universitario de Estudios Avanzados, Universidad de La Laguna, La Laguna 38203, Spain}

\author{M. B. Plenio}
\affiliation{Institut f\"ur Theoretische Physik, Albert-Einstein-Allee 11, Universit\"at Ulm, D-89069 Ulm, Germany}

\author{A. del Campo}
\affiliation{Theoretical Division,  Los Alamos National Laboratory, Los Alamos, NM 87545, USA}
\affiliation{Center for Nonlinear Studies,  Los Alamos National Laboratory, Los Alamos, NM 87545, USA}

\begin{abstract}
We explore heat transport across an ion Coulomb crystal beyond the harmonic regime by tuning it
across the structural phase transition between the linear and zigzag configurations. This demonstrates
that the control of the spatial ion distribution by varying the trapping frequencies renders ion
Coulomb crystals an ideal test-bed to study heat transport properties in finite open system of tunable
non-linearities.
%We analyze the heat transport across a trapped ion chain driven by counterpropagating laser beams acting on
%the ions on the edges. The control of the spatial ion distribution by varying the trapping frequencies makes
%ion chains an ideal test-bed to study heat transport properties in finite open system of tunable low dimensionality.
%We explore heat transport across a structural phase transition between the linear and zigzag configurations,
%identifying the condition for optimal heat transport.
\end{abstract}

%05.60.-k:	Transport processes
%05.60.Cd:	Classical transport
%64.60.Ht: Dynamic critical phenomena
%05.70.Fh: Phase transitions: general studies
%37.10.Ty : ion trapping

\maketitle

%\section{Introduction}

Ultracold ion Coulomb crystals represent one of the most promising platforms for the
simulation of many-body physics thanks to the high degree of spatial and temporal
control of mesoscopic ion crystals they afford us with \cite{HRB08,Schaetz12,BR12}.
Recent years have seen a shift away from the study of ground and thermal state properties,
towards the exploration of the potential role of ion traps as a test-bed for models
of non-equilibrium statistical mechanics.
%
% I have commented those out as in my view they break the flow at this stage, perhaps one can mention
% them a little later?
% Indeed, trapped ions have been applied to the study of static and dynamic friction in cold atom tribology
% \cite{Benassi11}, pattern formation \cite{LC11}, and non-equilibrium thermodynamics \cite{Huber08, Abah12}.
%
%Trapped-ion chains also offer the possibility of testing critical behavior.

In this context it is important to recognize that in addition to the electronic spin degrees of
freedom trapped ions also possess motional degrees of freedom that can exhibit highly non-trivial
static and dynamical properties including classical and quantum phase transitions. Indeed, ion
Coulomb crystals confined in ion traps may support a wide variety of phases including a linear
chain and a doubly-degenerate zigzag phase, extending further to increasingly complex configurations
in two and three spatial dimensions \cite{Walther92a,Walther92b,Dubin99}. The associated structural
phase transitions between those configurations are generally of first order, with the exception of
the linear-to-zigzag phase transition which is known to be of second order \cite{Schiffer93,Piacente04,Fishman08}.
As a symmetry breaking scenario, it provides a natural testing ground for universal dynamics of phase
transitions and topological defect formation \cite{Fishman08,Retzker08,delCampo10,DeChiara10}, recently 
explored in the laboratory  \cite{Schaetz13,EH13,Ulm13,Tanja13}.
%. The formation
%of defects upon traversal of this phase transition at a finite rate due to quenches of an external control
%parameter follows a universal scaling law which has been observed in the laboratory \cite{Schaetz13,EH13,Ulm13,Tanja13}.

Another fundamental setting in non-equilibrium statistical mechanics considers the thermal transport in
low-dimensional systems, which exhibits a rich variety of anomalous features, including the breakdown
of Fourier's law of heat conduction, instances in which sub-diffusive and super-diffusive behavior
can be observed \cite{LWH02,Alonso02,Alonso04}, as well as the divergence of the thermal conductivity
with the system size \cite{LLP03,Dhar08}. Most rigorous theoretical results have been obtained for exactly
solvable quasi-free models while systems with non-linearities are typically exceedingly difficult to treat.
Equally, the controlled generation of non-linear physics in mesoscopic ion crystals is non-trivial and much
recent progress has concerned harmonic models of complex networks and trapped-ion chains \cite{MP12,Bermudez13,Ramm13}.
The richest phenomenology however can be expected in non-integrable models \cite{HLZ} which mandates
the development of both theoretical and experimental methods for their examination.

%Further experimental results measuring the transport of energy from a pulsed excitation applied onto a
%single ion in the chain have been recently reported \cite{Ramm13}.

%In this Letter, we consider continuously driven ion chains between two thermal reservoirs as a platform in
%which to explore heat transport across low-dimensional systems that experience a structural phase transition.
%As the ion chain crosses the transition from the linear to zigzag phase, the system may exhibit
%both significant non-linearities and axial-transverse mode coupling which can lead to qualitative changes
%in both the local temperature profile and the total heat flux through the chain. This advances further the
%case for trapped ions as a model system for the examination of challenging problems in mesoscopic physics.

In this Letter, we advance further the case for trapped ions as a model system for the examination of challenging
problems in mesoscopic physics by considering continuously driven ion chains between two thermal reservoirs as
a platform in which to explore heat transport across an ion Coulomb crystal that experience a structural phase
transition. As the ion chain crosses the transition from the linear to zigzag phase, the system may exhibit
both significant non-linearities and axial-transverse mode coupling which can lead to qualitative changes
in both the local temperature profile and the total heat flux through the chain that may be observed experimentally.

{\it The system dynamics.-}
We consider an effectively $2D$-dimensional system composed of $N$ ions of mass $m$, charge $Q$, positions ${\bf q}_n=(q_{x,n},q_{y,n})$ and momenta ${\bf p}_n=(p_{x,n},p_{y,n})$, with $n=1,\dots,N$,
which are confined in a trap with axial frequency $\nu$ along the $x-\,$axis and transversal frequency $\nu_t$
along the $y-\,$axis. The Hamiltonian of the system can be written as
\begin{equation}
    H=\frac{1}{2m}\sum_{n=1}^{N}\left(p_{x,n}^2+p_{y,n}^2\right)+{\cal V}\, .
\end{equation}
The interaction potential ${\cal V}$ accounts
for both the harmonic trap and the Coulomb repulsion, and is given by
\begin{equation}\label{pot}
    {\cal V}=\frac{m}{2}\sum_{n=1}^{N}\left(\nu^2q_{x,n}^2+\nu_t^2q_{y,n}^2\right)+\frac{1}{2}\left(\frac{Q^2}{4\pi\varepsilon_0}\right)\sum_{n=1}^{N}\sum_{l\ne n}^{N}
    \frac{1}{|{\bf q}_n-{\bf q}_l|}\,.
\end{equation}
A quasi-linear confinement of the ions along the $x-\,$axis can be achieved by considering a strongly
anisotropic trap, with $\nu_t\gg\nu$.  We note that at variance with lattice systems, none of the ions
is pinned, which allows for an intricate interplay between axial and radial modes of motion.

We assume that the dynamics due to the external Doppler cooling lasers acting on the ions can be
modeled as Langevin thermostats. This together with the typical separations between the ions (generally
of the order of microns) justifies an intrinsically noisy classical description of the dynamics,
%given by the $4N-\,$ equations of motion
%
\begin{eqnarray}\label{eqn2}
    &&dq_{\mu,n}=\frac{p_{\mu,n}}{m}\,dt \\
    &&dp_{\mu,n}=-\left(\frac{\partial {\cal V}}{\,\,\partial q_{\mu,n}}+\frac{\eta_{\mu,n}}{m}\,p_{\mu,n}\right)dt+\sqrt{2\,D_{\mu,n}}\,\,dW_{\mu,n}\,, \nonumber
\end{eqnarray}
where $\eta_{\mu,n}$ and $D_{\mu,n}$ are the friction and diffusion coefficients, respectively,
$dW_{\mu,n}$ denote the Wiener processes resulting from the Gaussian white noise forces $\varepsilon_{\mu,n}(t)$
associated with the diffusion induced by the interaction with the laser beams, which satisfy
$\la\varepsilon_{\mu,n}(t)\ra\,=\,0 $ and $\la\varepsilon_{\mu,n}(t)\,\varepsilon_{\mu,n}(t^\prime)\ra\,=\,2\,D_{\mu,n}\,\delta(t-t^\prime)$,
and $\mu=(x,y)$.

For small laser intensities the friction and diffusion coefficients can be obtained from the Doppler
cooling expressions 
\begin{eqnarray}\label{dos}
    \eta_{\mu,n}&=&-4\hbar k_{\mu,n}^2\left(\frac{I_{\mu,n}}{I_0}\right)\frac{\left(2\delta_{\mu,n}/\Gamma\right)}{\left[1+4\delta_{\mu,n}^2/\Gamma^2\right]^2},\nonumber\\
    D_{\mu,n}&=&\hbar^2 k_{\mu,n}^2\left(\frac{I_{\mu,n}}{I_0}\right)\frac{\Gamma}{\left[1+4\delta_{\mu,n}^2/\Gamma^2\right]}\,,
\end{eqnarray}
where $I_{\mu,n}/I_0$ is the normalized intensity of the laser beam acting on the $n$-ion along the
$\mu$-direction, $k_{\mu,n}$ is the corresponding laser wavelength, $\delta_{\mu,n}=\omega_{\mu,n}-\omega_0$
is the detuning of the laser frequency $\omega_{\mu,n}$ with respect to the frequency $\omega_0$ of a
selected atomic transition in the ions, and $\Gamma$ is the natural linewidth of the excited state in
such transition \cite{Phillips92}.
%For a given atomic transition, we will assume the coefficients $\eta_{\mu,n}$ and
%$D_{\mu,n}$ as functions of the detuning and the normalized intensity of the laser beams.

\begin{figure}
\centering\includegraphics[width=1\linewidth]{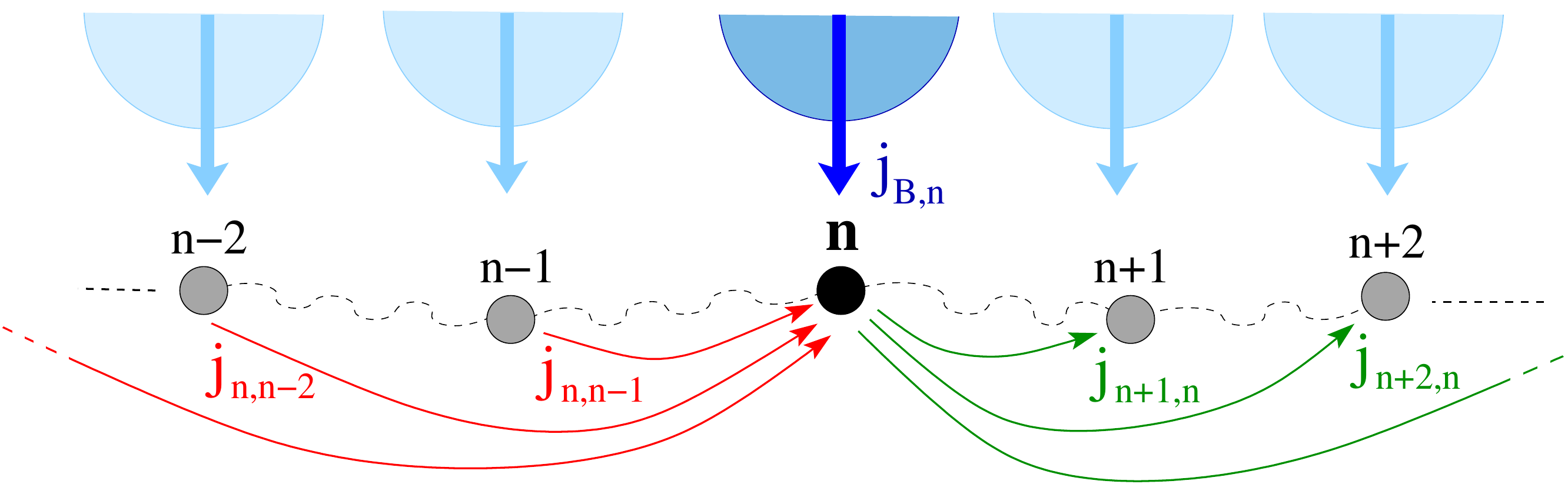}
\caption{(Color online) An illustration of some of the energy currents associated with the $n$-ion of the chain.
}
\label{fig_current}
\end{figure}

{\it Heat flux and local kinetic temperature.-}
A discrete definition of the heat current through the chain can be obtained from the local energy density associated with each ion \cite{LLP03,Dhar08}, which can be written as
\begin{equation}\label{hn}
    h_n=\frac{1}{2m}\left(p_{x,n}^2+p_{y,n}^2\right)+V\left({\bf q}_n\right)+\frac{1}{2}
    \sum_{l\ne n}^N U\left(|{\bf q}_l-{\bf q}_n|\right)\,,
\end{equation}
where $V$ represents the harmonic trap and $U$ the Coulomb term of the interaction potential
${\cal V}$ given in Eq. (\ref{pot}). The time derivative of $h_n$ leads to the discrete
continuity equations
\begin{equation}\label{conti1}
    \frac{dh_n}{dt}=\sum_{l<n}^N j_{n,l} - \sum_{l>n}^N j_{l,n}+j_{B,n}\,,
\end{equation}
where
\begin{equation}
    j_{n,l} = -\frac{1}{2m}\sum_{\mu=\{x,y\}}\frac{\partial U \left(|{\bf q}_l-{\bf q}_n|\right)}{\partial q_{\mu,n}}\,\left(p_{\mu,n}+p_{\mu,l}\right)
\end{equation}
can be identified as the energy current from the $l$-ion to the $n$-ion. For the $n$-ion,
the first term in Eq. (\ref{conti1}) corresponds to the total energy current coming from the ions on the
left, whereas the second term is the total energy current going to the ions on the right, see
Fig. (\ref{fig_current}).
%%%%%%%%%%%%%%%%%%%%%%%%%%%%%%%%%%%%%%%%%%%%%%%%%
%%%%%%%%%%%%%%%%%%%%%%%%%%%%%%%%%%%%%%%%%%%%%%%%%
The  last term
\begin{equation}
    j_{B,n}=\sum_{\mu=\{x,y\}}\frac{p_{\mu,n}}{m}\left(\,-\frac{\eta_{\mu,n}}{m}\,p_{\mu,n}+\varepsilon_{\mu,n}\,\right)
\end{equation}
is the energy current from the laser reservoirs.

The steady-state average of Eq. (\ref{conti1}) implies the balance
%$\sum_{l<n} (\la j_{n,l}\ra\,+\,\la j_{B,n}\ra)\,=\,\sum_{l>n} \la j_{l,n}\ra$,
%
\begin{equation}\label{steady1}
\sum_{l<n}^N\la j_{n,l}\ra\,+\,\la j_{B,n}\ra\,=\,\sum_{l>n}^N \la j_{l,n}\ra\,,
\end{equation}
%
%which indicates a balance
between the average rate at which each ion receives energy from the ions on
the left and the laser beams, and the average rate at which such ion transfers energy to the ions on the
right. The average of the energy currents from the reservoirs can be obtained using Novikov's theorem
\cite{Novikov},
\begin{equation}
    \la j_{B,n}\ra\,=\,\frac{1}{\,m^ 2}\,\sum_{\mu=\{x,y\}}\left(-\eta_{\mu,n}\la p_{\mu,n}^2\ra+m\,D_{\mu,n}\right)\,.
\end{equation}

The total heat current can be derived from a discrete description of the continuity equation \cite{LLP03,Dhar08}
\begin{equation}\label{conti}
    \frac{\,\partial}{\partial t}h({\bf q},t)+\nabla\cdot {\bf j}({\bf q},t)=\sum_{n=1}^N j_{B,n}(t)\delta({\bf q}-{\bf q}_n)\,,
\end{equation}
by taking the energy and heat flux densities as
$h({\bf q},t)=\sum_{n=1}^Nh_n(t)\delta({\bf q}-{\bf q}_n)\,$,
${\bf j}({\bf q},t)=\sum_{n=1}^N{\bf j}_n(t)\delta({\bf q}-{\bf q}_n)\,,$
%
%\beqa
%h({\bf q},t)&=&\sum_{n=1}^Nh_n(t)\delta({\bf q}-{\bf q}_n)\,, \\ \nonumber
%{\bf j}({\bf q},t)&=&\sum_{n=1}^N{\bf j}_n(t)\delta({\bf q}-{\bf q}_n)\,,
%\eeqa
respectively, with $h_n$ being the local energy density defined in Eq. (\ref{hn}) and ${\bf j}_n$ the local flux.
A Fourier analysis of Eq. (\ref{conti}) leads to ${\bf j}_n(t)={\bf q}_n\left(\frac{dh_n}{dt}-j_{B,n}\right)+h_n\frac{d{\bf q}_n}{dt}\,.$
%
%\begin{equation}
%    {\bf j}_n(t)={\bf q}_n\left(\frac{dh_n}{dt}-j_{B,n}\right)+h_n\frac{d{\bf q}_n}{dt}\,.
%\end{equation}
%
Then the total heat flux, obtained by integration of the flux density over the chain volume, reads
\beqa\label{jt}
{\bf J}(t)
%&=&\int {\bf j}({\bf q},t)\,d{\bf q}=\sum_{n=1}^N{\bf j}_n(t)\nonumber\\
%&=&
=\frac{1}{m}\sum_{n=1}^Nh_n{\bf p}_n+\sum_{n=1}^{N-1}\sum_{l=1}^n({\bf q}_{n+1}-{\bf q}_l)j_{n+1,l}\,.
\eeqa
In the steady state, the averaged total heat flux is determined by just the local fluxes coming from the reservoirs,
and applying  Novikov's theorem \cite{Novikov} it follows that
\begin{eqnarray}\label{steady3}
    \la{\bf J}\ra&=&-\sum_{n=1}^N\la{\bf q}_nj_{B,n}\ra\nonumber\\
    &=&\frac{1}{\,m^2}\,\sum_{n=1}^N\,\sum_{\mu=\{x,y\}}\left(\,\eta_{\mu,n}\la p_{\mu,n}^2{\bf q}_n\ra-m\,D_{\mu,n}\,\la{\bf q}_n\ra\,\right)\,.
\end{eqnarray}

A discrete approach can also be considered to define a local temperature through the chain in terms of dynamical
variables. According to the virial theorem, the local kinetic temperature $T_n$ of each ion can be defined
from its kinetic energy
\begin{equation}
   T_n\,=\,\frac{1}{2m}\sum_{\mu=\{x,y\}}\la\,p_{\mu,n}^2\,\ra_{\varepsilon}
\end{equation}
where $\la\,\cdot\,\ra_\varepsilon$ indicates the average over an ensemble of stochastic trajectories.

%%%%%%%%%%%%%%%%%%%%%%%%%%%%%%%%%%%%%%%%%%%%

{\it Numerical experiments.-}
We consider a chain composed of $N$ ions and analyze the response of the local temperature and the total heat
flux to the phase transition from a quasi-linear to the planar zigzag spatial configuration as the transversal
frequency of the trap is lowered.
% To gain some intuition we notice that to leading order in an expansion of the potential ${\cal V}$ around the
% equilibrium positions of the ions, the system is equivalent to a 1D chain with an  on-site potential including
% both a quadratic and a quartic term, and a bilinear coupling between nearest-neighboring sites \cite{Fishman08}.
% The quadratic coefficient is proportional to $(\nu_t^2-\nu_{tc}^2)$ changing sign as the transverse frequency
% $\nu_t$ is decreased beyond the critical value $\nu_{tc}$. Consequently, the onsite potential can vary from a
% single well to a double well. However, this expansion is only valid in the neighborhood of the transition, and
% we will study heat conduction  beyond its regime of applicability.
We consider a chain of $^{24}$Mg$^+$ ions, with $N=30$, and fix an axial frequency of the trap to
$\nu=2\pi\times 50\,$kHz. We study the dynamics for different transversal frequencies $\nu_t=\alpha\nu$.

An analysis of the static properties of the ion chain in the thermodynamic limit provides an  estimate of the
local value of the critical ratio of the trap frequencies leading to the phase transition between the linear and
the zigzag configurations \cite{Fishman08},
$\alpha_c(x)=\sqrt{\frac{7\zeta(3)}{2M\nu^2}\left(\frac{Q^2}{4\pi\varepsilon_0}\right)\,}\,[n(x)]^{3/2}$, where
$\zeta$ is the Riemann-zeta function, $n(x)=(3N/4L)[1-(x/L)^2]$ is the equilibrium linear density of ions along the trap axis
as a function of  distance $x$ from the chain's center, and the half-length of the chain $L$
\cite{Dubin97}. Due to the axial harmonic confinement the center of the chain experiences a higher axial density
and Coulomb repulsion,  making the phase transition spatially inhomogeneous.
%Therefore, $\alpha_c(x)$ displays a smooth dependence on $x$, increasing toward the center
%of the chain so that ions located in the center of the chain cross the transition to the zigzag configuration first.

In our numerical studies we assume that the ions are initially at rest and arranged with random positions in the
close vicinity of the linear configuration. Then the reservoir lasers that act on the selected ions are switched
on instantaneously. To determine the friction and the diffusion coefficients that characterize the interaction of
the $^{24}$Mg$^+$ ions with the laser beams, we have considered the Doppler cooling expressions (\ref{dos})
applied to the atomic transition $3s^2S_{1/2} \longrightarrow 3p^2P_{1/2}$ with frequency $\omega_0=2\pi\times1069\,$THz
\cite{NIST} and an excited state  natural linewidth $\Gamma=2\pi\times41.296\,$MHz \cite{Ansbacher89}. Given these
values, the interaction
of a laser beam with an ion is a function of the normalized intensity $(I_{\mu,n}/I_0)$ and the detuning
$\delta_{\mu,n}$ of the laser beams. Note that in order to avoid excessively small time steps the numerical model
neglects micromotion. While this represents a significant approximation in the zigzag configuration of an ion
Coulomb crystal in a rf-Paul trap, it should be noted, however, that micromotion is absent in Penning traps in
which analogous structural phase transitions were recently observed and Doppler cooling can be implemented \cite{Segal13}.

To drive a heat current through the chain, the ions at opposite ends of the ion crystal are subjected to
different laser beams. In particular, we consider that the three leftmost (rightmost) ions interact with
laser beams with normalized intensity $I_L=(I_{\mu,n}/I_0)=0.08$ ($I_R=I_L$) and detuning $\delta_L =
\delta_{\mu,n}=-0.02\Gamma$ ($\delta_R=\delta_{\mu,n}=-0.1\Gamma$), with $n=1,2,3$ ($n=N-2,N-1,N$). The
different detunings $\delta_L$ and $\delta_R$ lead to effective reservoirs operating at different temperatures
on both ends of the chain, and therefore a stationary non-equilibrium heat current. We assume that no laser
beams are acting on the inner ions, corresponding to $n=4,\dots,N-3$. If local addressing of individual ions
is not available, the use of $^{25}$Mg$^+$ ions at the opposite ends of the chain that is otherwise composed
of $^{24}$Mg$^+$ can allow for highly localized heat baths by means of frequency selection. $^{25}$Mg$^+$ ions
that are not part of the zigzag structure will not affect the essential features of the structural phase transition.
We are principally concerned with the steady state behaviour of the system. As a criterion to determine that the
system has reached the steady state we verify expression (\ref{steady1}) for each ion, and also apply equality
(\ref{steady3}) to the whole chain (see Supplementary Material for more details).

As Fig. (\ref{fig_temp}) shows,
the temperature profile adopts a gradient along the chain which progressively increases as the transverse
frequency of the trap is reduced to drive the transition from the linear to the zigzag spatial distributions.
The numerical simulations indicate that the phase transition first emerges for $\alpha_c(0)\simeq 11.6$.
%%%%%%%%%%%%%%%%%%%%%%%%%%%%%%%%%%%%%%%%%%%%%%%%%
\begin{figure}[t]\centering
\includegraphics[width=0.49\linewidth]{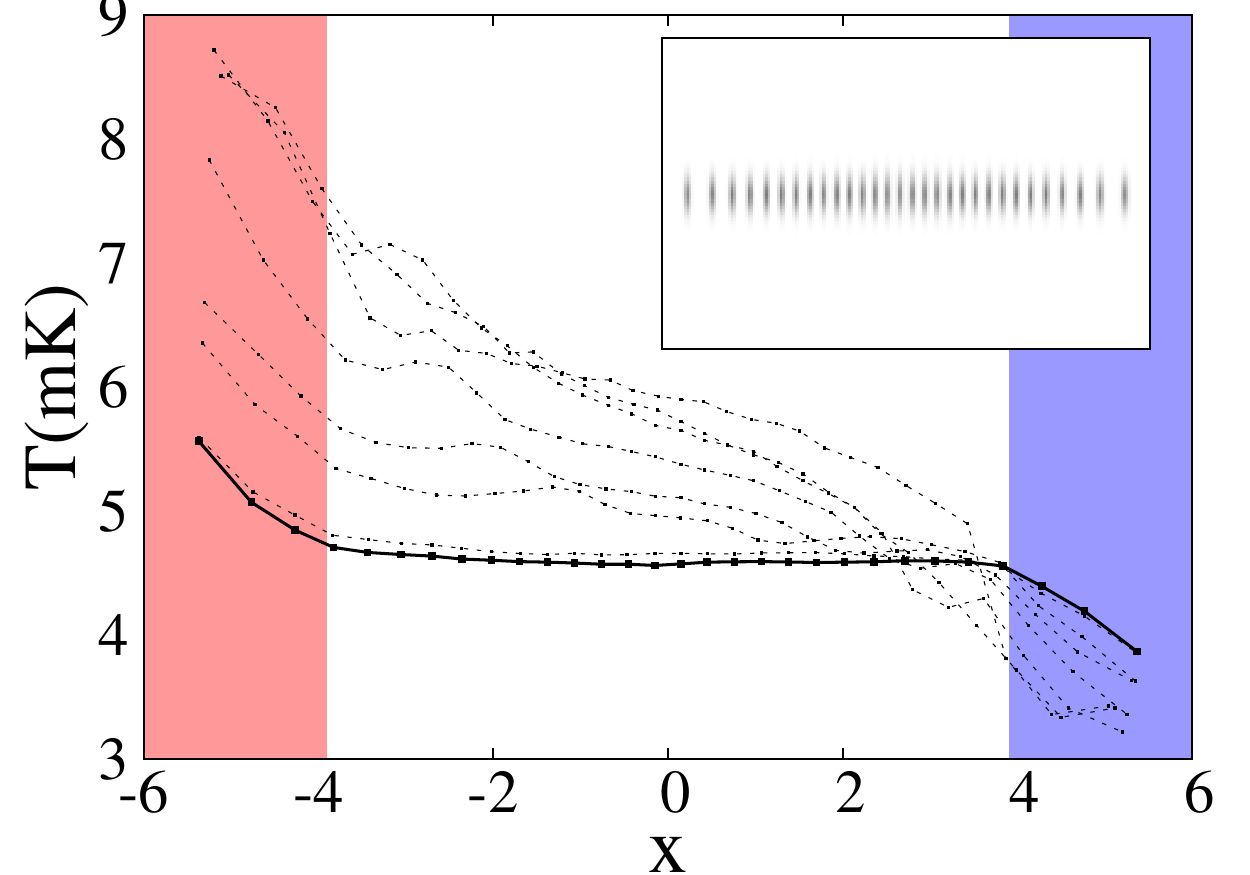}
\includegraphics[width=0.49\linewidth]{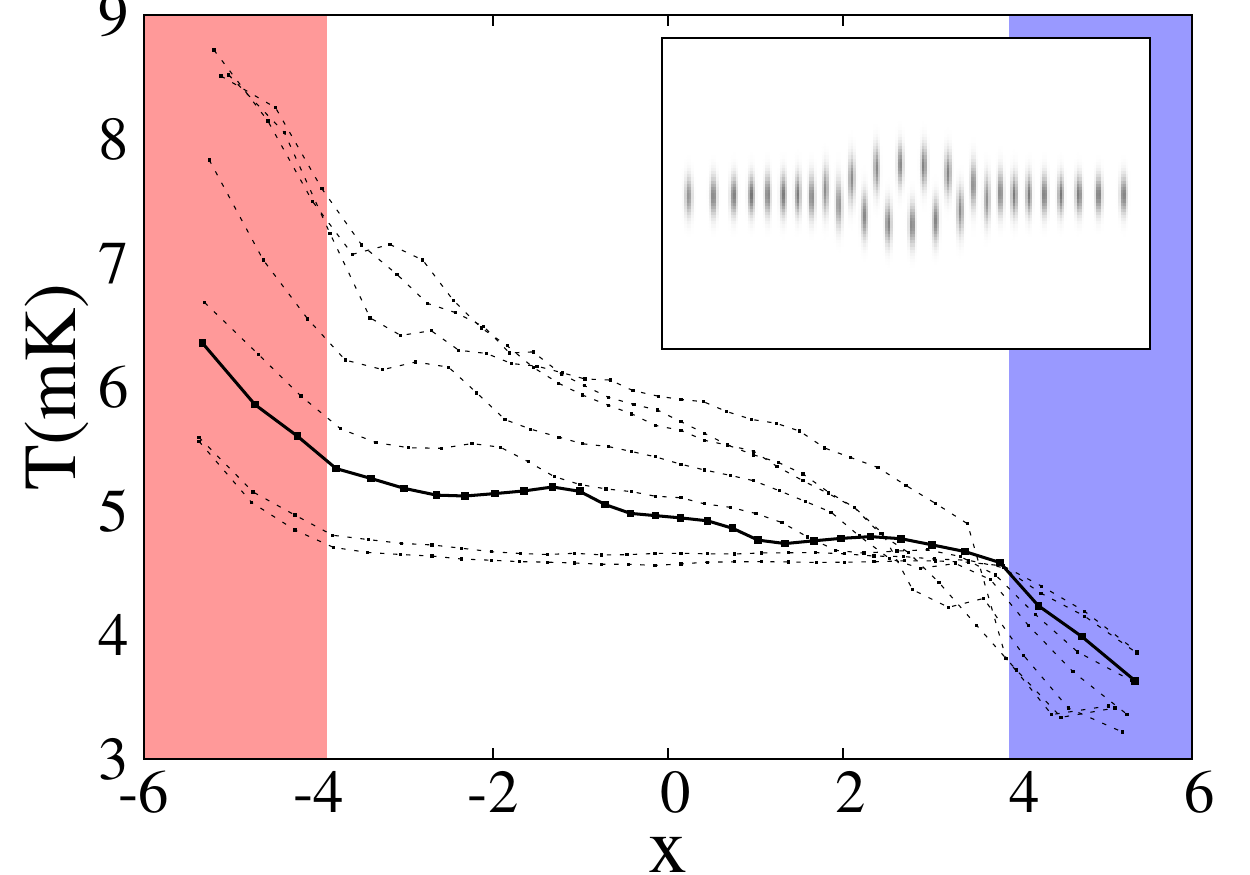}
\includegraphics[width=0.49\linewidth]{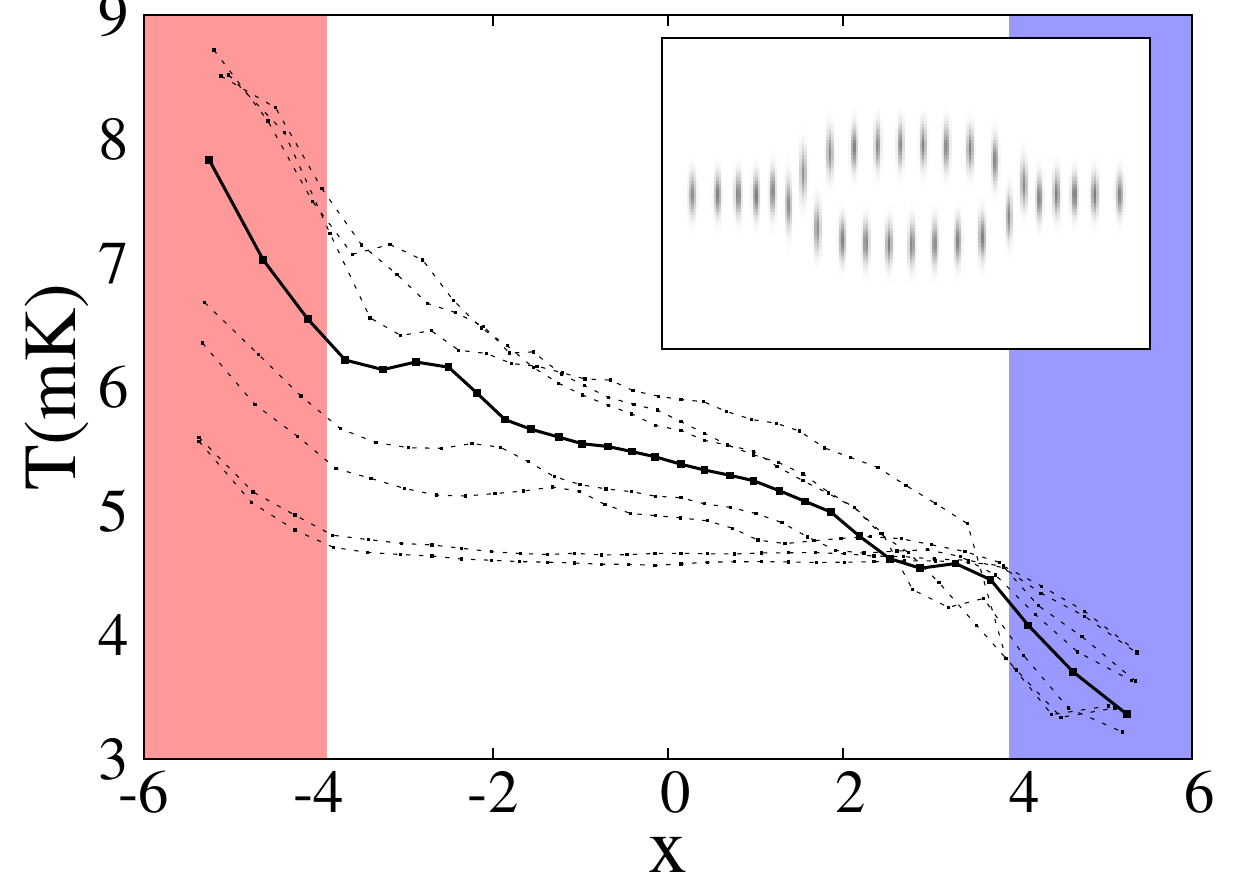}
\includegraphics[width=0.49\linewidth]{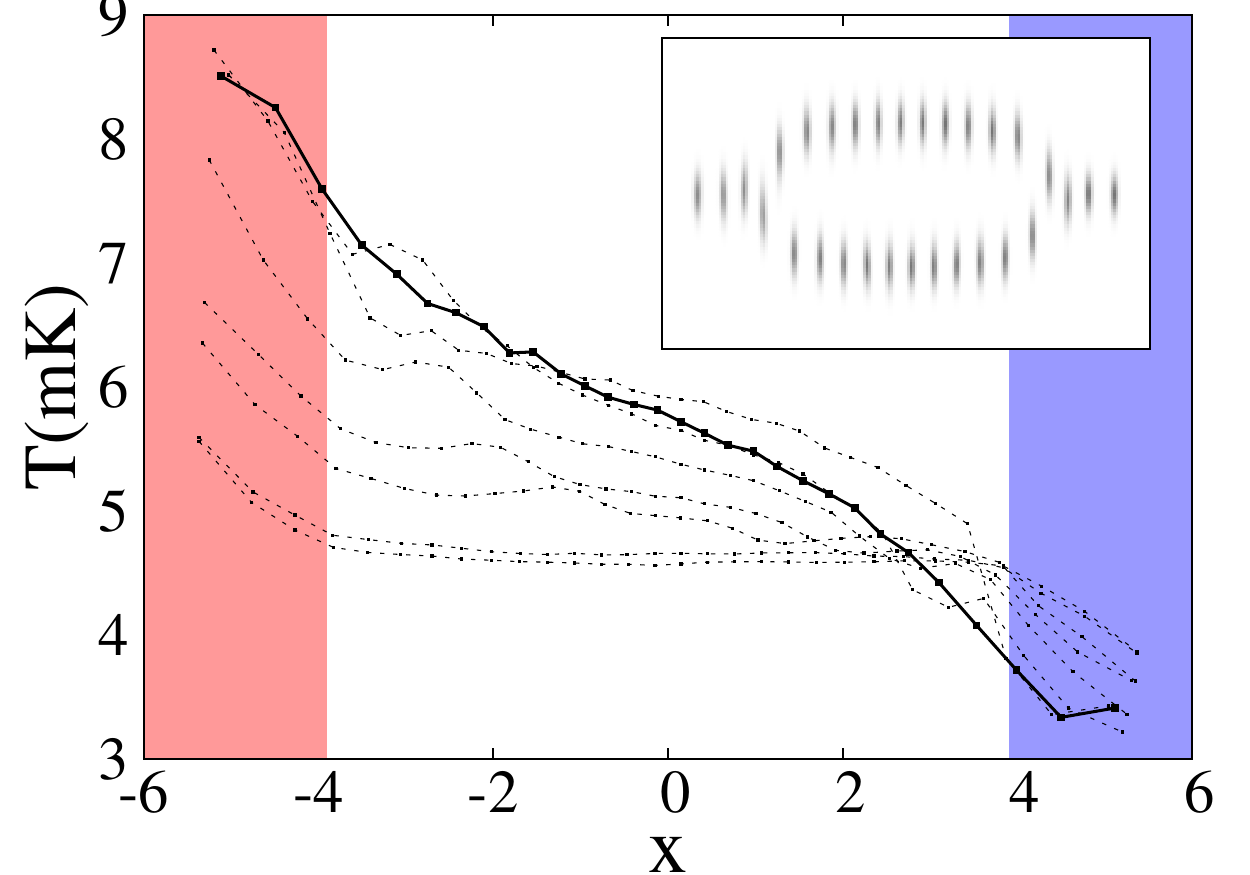}
\caption{(Color online) Local temperature profiles.
The dashed lines in the four panels indicate the profile of the local temperatures along the chain vs the distance $x$ from the chain's center, for a ratio of the trap frequencies that ranges from $\alpha=13$ to $\alpha=7$. In the upper panels the thick lines correspond to $\alpha=13\,$(left) and $\alpha=11\,$(right), and in the lower panels to $\alpha=9\,$(left) and $\alpha=7\,$(right). The lines between points are drawn to guide the eye. The insets show the steady ion distribution obtained from a single stochastic trajectory for a chain with the value of $\alpha$ depicted with the thick line.
The color boxes indicate the ions that are connected to the laser beams.
}
\label{fig_temp}
\end{figure}
%%%%%%%%%%%%%%%%%%%%%%%%%%%%%%%%%%%%%%%%%%%%%%%%%
For $\alpha>\alpha_c(0)$ the chain is fully linear and the temperatures $T_n\,(n=4,\dots,N-3)$ of the inner
ions tend to settle on a constant value for all $n$,  which for this system is close to the mean temperature
$(T_1+T_N)/2$ that is expected in a homogeneous harmonic chain with nearest-neighbor interactions \cite{RLL67},
despite the presence of the axial quadratic potential \cite{PC05}.
Indeed, a spectral analysis of the steady evolution of the coordinates of the inner ions in linear chains indicates
that their axial dynamics is close to the Brownian motion of a simple harmonic oscillator with characteristic frequency
$\nu$.  Figure (\ref{fig_spec}) illustrates that the spectra of these ions are dominated by a well defined single peak
corresponding to the axial trap frequency. Also an ordered series of much less intense peaks corresponding to higher
order multiples of $\nu$ can be distinguished. The very low intensity of the transversal spectrum evidences the minor role of this mode, due to the strong trap confinement in radial direction.

The simple axial spectrum corroborates the expectation that a harmonic approximation to the system
Hamiltonian is valid; the system is effectively integrable, and the heat carriers are freely propagating
phonons. The lack of temperature gradient observed in linear chains is characteristic of this ballistic behavior.

This situation is expected to change when the chain approaches the structural phase transition at $\alpha_c(0)$ where
the chain buckles with the growth of the zig-zag soft mode. Near this point non-linearities as well as mode coupling
between axial and radial modes are expected to play an increasing role and the harmonic chain description is expected
to fail. While the non-linearities lead to scattering, the coupling between axial and radial modes lead to an effective
dynamics akin to dephasing noise. Both effects, if significant, are known to contribute to the formation of a
temperature gradient in the chain. It should be noted that non-linearities tend to be relatively small unless the
chain is very close to the phase transition point \cite{Marquet02}. Hence we expect coupling between radial and axial modes to dominate.
The deviations from the harmonic picture are also witnessed by the spectra of the axial motion of the
ions in
%the broken-symmetry phase of the transition characterized by
the zig-zag configuration, which exhibit more complex
features that were absent in the linear chains, see Figure (\ref{fig_spec}), and that can be assigned to the coupling
with the transverse motion.
The presence of bands of irregular patterns extend across the axial and transverse spectra suggests underlying domains
of chaotic dynamics arising from such coupling. As the harmonic chain description ceases to be valid, the heat transport
through the chain is modified, as evidenced by the emergence of a temperature gradient, signalling a possible diffusive
behavior of the heat carriers.
%%%%%%%%%%%%%%%%%%%%%%%%%%%%%%%%%%%%%%%%%%%%%%%%%%%%%%%%%%%
\begin{figure}[t]\centering
\includegraphics[width=0.98\linewidth]{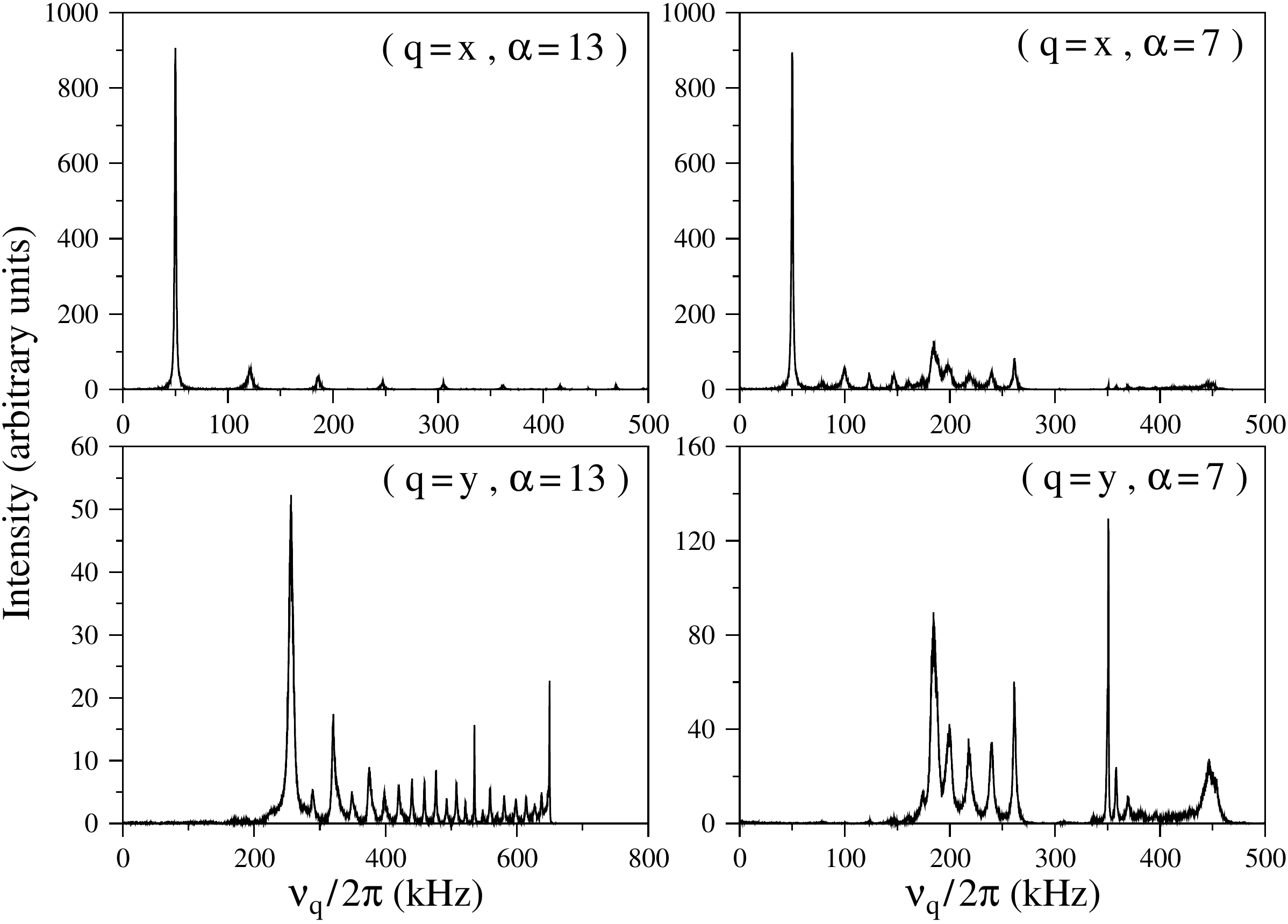}
\caption{
Spectra of the axial and the transversal motions for the central ion in a linear chain $(\alpha=13)$ and in a chain with zig-zag spatial distribution $(\alpha=7)$, see Fig. (\ref{fig_temp}). The spectra have been obtained from the Fast Fourier Transform (FFT) of the steady evolution of the axial $(q=x)$ and transversal $(q=y)$ coordinates.
}
\label{fig_spec}
\end{figure}
%%%%%%%%%%%%%%%%%%%%%%%%%%%%%%%%%%%%%%%%%%%%%%%%%%%%%%%%%%%%
%The discrepancy between the steady temperatures of the ions that are connected to the laser beams and the respective limit temperatures in the Doppler cooling of the isolated ions, $T_L>T_n\,(n=1,2,3)$ and $T_R<T_n\,(n=N-2,N-1,N)$, indicates that the dissipation rates due to the interactions with the lasers are not strong enough to overcome the effect of the trap and Coulomb interactions. Time scales of the heat reservoirs short enough to really set the temperature of the ions on the edges to $T_L$ and $T_R$ could be achieved by more intense lasers.
%\section{Variation of heat flux across the phase transition}

\begin{figure}[h]\centering
\includegraphics[width=0.8\linewidth]{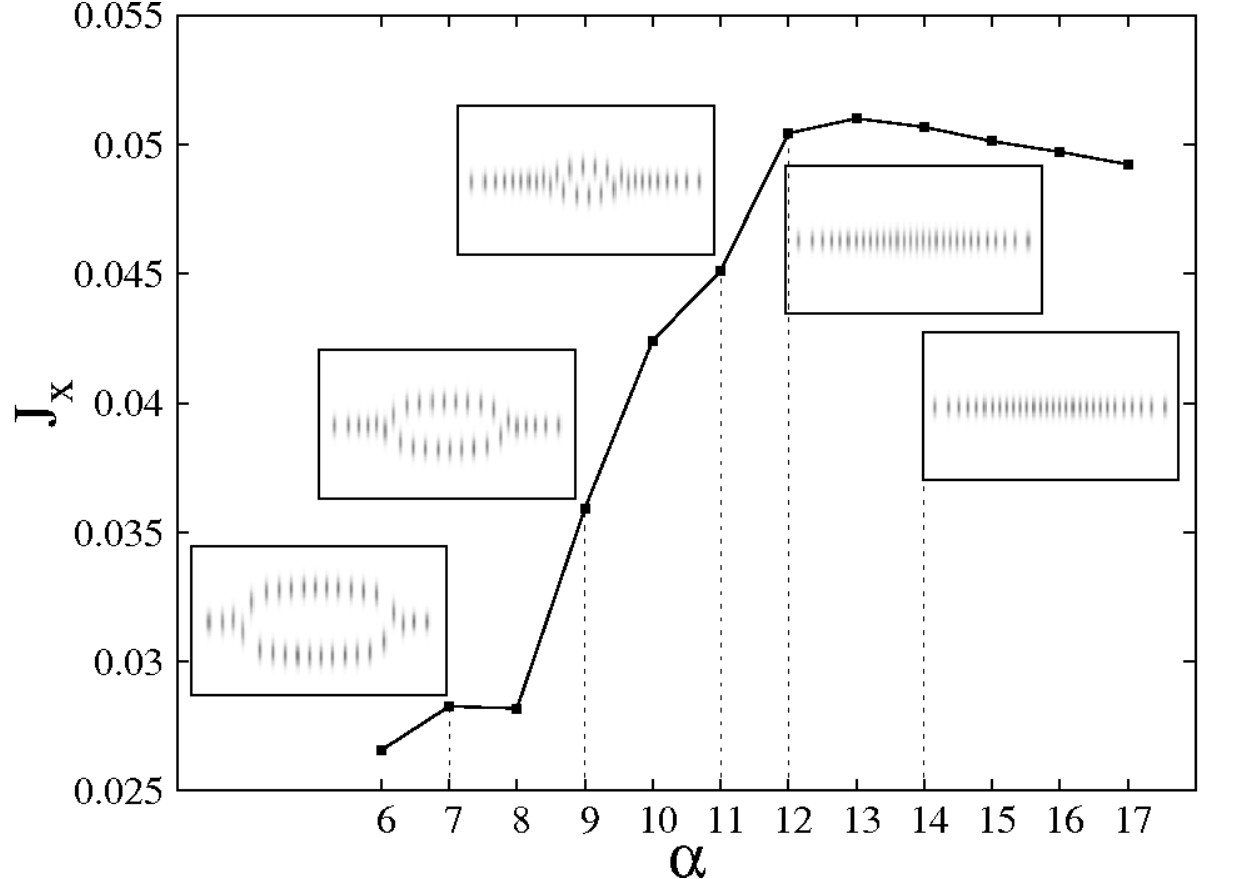}
\caption{
Total heat flux in the axial direction as a function of the trap frequency ratio $\alpha$.
%The solid line is drawn to guide the eye. The vertical dashed lines indicate the values of $\alpha$ selected to depict the steady ion distribution obtained from a single stochastic trajectory.
The maximum heat transport is achieved in the high-symmetry (linear) phase in the proximity of the critical point, with $\alpha\approx 13$.
}
\label{fig_flux}
\end{figure}

Further to the emergence of Fourier's law in the ion chain when approaching the structural phase transition,
one may also observe clear signatures of the structural phase transition in the heat flux across the chain
as demonstrated by figure (\ref{fig_flux}).
%commented? in which a clear sensitivity of the total heat flux along the axial
%direction on the crossing of the phase transition between the linear and zig-zag configurations can be observed.
%%%%%%%%%%%%%%%%%%%%%%%%%%%%%%%%%%%%%%%%%%%%%%%%%%%%%%%%%%%%55
%%%%%%%%%%%%%%%%%%%%%%%%%%%%%%%%%%%%%%%%%%%%%%%%%%%%%%%%%%%%%%%%%%%%%%
Initially, on approaching the structural phase transition from the linear chain, one observes a small increase
of the heat flux. This has two origins. Firstly, in the proximity of the transition the transverse modes will
start to contribute to transport. Secondly, the increased thermal motion of the ions as the chain softens upon
approach of the phase transition lead to an increased level of fluctuations. This noise may assist transport as it
overcome the effects of spatial inhomogeneity in the chain \cite{PlenioH08}. Upon further decrease of $\alpha$ the
chain buckles, leading to two possible heat conduction paths, while the inter-ion distances increase which in turn
leads to a reduction in the interaction between neighboring ions and therefore a reduction of the heat flux.
While initially, close to the phase transition, the increase in distance is compensated for by the formation of
two independent channels, deeper in the zigzag configuration this is not the case anymore and the heat flux
reduces.

Before closing we point out that a thermal conductance could be estimated from the temperature gradient and
the axial heat flux measured in a non-equilibrium steady state. The spatial constrains imposed by the finite
size and low-dimensionality of trapped-ion chains prompt the analysis of the thermal conductance as a function
of the length, instead of a size-independent thermal conductivity of interest in a macroscopic model of thermal
conduction.
An alternative approach to obtain the heat conductance of a specific ion chain could be based on Green-Kubo
type linear response expressions valid for the heat current in finite low-dimensional systems coupled to heat
reservoirs \cite{Dhar08,Kundu09}.

{\it Conclusions.-}
Our analysis based on the local temperature profile and the total heat flux indicates that trapped-ion chains
exhibit anomalous heat transport. The linearly distributed ions resemble harmonic chains, and therefore an
integrable system, in which the free energy transport along the chain by non-interacting axial modes precludes
the establishment of a temperature gradient and would lead to a divergent thermal conductivity.
The phase transition from the linear to the bi-dimensional zigzag configuration induces a coupling between axial
and transverse modes that hinders the transport of energy along the chain, and allows the emergence of a central
domain in which a temperature gradient can be set up. Such domain grows as the transversal frequency is lowered
and the bidimensional configuration extends towards the ends of the chain, resulting in a significant decrease
of the axial heat flux. Heat transport is optimal in the linear configuration in the proximity of the critical
point.

{\it Note:} After the completion of this work, we learned about reference \cite{Freitas13} devoted to the study
of quantum heat conduction in harmonic ion chains.
%%%%%%%%%%%%%%%%%%%%%%%%%%%%%%%%%%%%%%%%%%%%%%%%%%%

{\it Acknowledgment.-}
It is a  pleasure to acknowledge discussions with T. E. Mehlst\"aubler,  J. M. Plata and D. Roy. This project was
funded by the Spanish MICINN, the European Union (FEDER) (FIS2010-19998), the EU Integrating project SIQS, the
EU STREP EQUAM, the Alexander von Humboldt Professorship (MBP) and the U.S. Department of Energy through the LANL/LDRD
Program and a LANL J. Robert Oppenheimer Fellowship (AdC).

%%%%%%%%%%%%%%%%%%%%%%%%%%%%%%%%%%%%%%%%%%%%%%%%%%

\newpage

%%%%%%%%%%%%%%%%%%%%%%%%%%%%%%%%%%%%%%%%%%%%%%%%%%%%%%%%%%%%%%%%%%%%%%

\appendix

\section{Dimensionless variables}

\vspace*{0.2cm}

The analysis of the system dynamics can be simplified considering the equations of motion (\ref{eqn2}) in terms of dimensionless variables. To define such variables we start by introducing a characteristic system length $\ell$, given by the relation
\begin{equation}
\ell^3=\frac{1}{m\nu^2}\left(\frac{Q^2}{4\pi\varepsilon_0}\right)\,.
\end{equation}
Defining the dimensionless ion coordinate and momentum vectors as
\begin{equation}
{\tilde {\bf q}}_n=({\tilde q}_{x,n}\,,\,{\tilde q}_{y,n})=\left(\,\frac{q_{x,n}}{\ell}\,,\,\frac{q_{y,n}}{\ell}\,\right)
\end{equation}
and
\begin{equation}
{\tilde {\bf p}}_n=(\tilde{p}_{x,n}\,,\,\tilde{p}_{y,n})=\left(\,\frac{p_{x,n}}{\ell m\nu}\,,\,\frac{p_{y,n}}{\ell m\nu}\,\right)\,,
\end{equation}
and the one-parameter dimensionless interaction potential
\begin{equation}\label{vad}
{\tilde {\cal V}}\,=\,\frac{{\cal V}}{\ell^2 m\nu^2}\,=\,
\frac{1}{2}\sum_{n=1}^{N}\left(\,{\tilde q}_{x,n}^2+\alpha^2\,{\tilde q}_{y,n}^2\,\right)
+\frac{1}{2}\sum_{n=1}^{N}\sum_{l\ne n}^{N}
\frac{1}{|{\tilde {\bf q}}_n-{\tilde {\bf q}}_l|}\,,
\end{equation}
where $\alpha=\nu_t/\nu$ is the aspect ratio of the trap frequencies, the equations of motions take the form
\begin{eqnarray}\label{eqn3}
&&d{\tilde q}_{\mu,n}=\tilde{p}_{\mu,n}\,d{\tilde t} \\
&&d{\tilde p}_{\mu,n}=-\left(\frac{\partial {\tilde {\cal V}}}{\,\partial {\tilde q}_{\mu,n}}+{\tilde \eta}_{\mu,n}\,\tilde{p}_{\mu,n}\right)d{\tilde t}+\sqrt{2\,{\tilde D}_{\mu,n}}\,\,d{\tilde W}_n\,, \nonumber
\end{eqnarray}
with the dimensionless time
\begin{equation}
{\tilde t}=\nu t,
\end{equation}
the Wiener processes
\begin{equation}
d{\tilde W}_{\mu,n}=\sqrt{\nu}\,dW_{\mu,n}\,,
\end{equation}
the friction coefficients
\begin{equation}
{\tilde \eta}_{\mu,n}=\frac{\eta_{\mu,n}}{m\nu}\,,
\end{equation}
and the diffusion coefficients
\begin{equation}
{\tilde D}_{\mu,n}=\frac{D_{\mu,n}}{\ell^2m^2\nu^3}\,.
\end{equation}

In terms of the dimensionless variables, the total heat flux (\ref{jt}) is expressed as
\begin{equation}
{\tilde {\bf J}}=\frac{{\bf J}}{\ell^3m\nu^3}=\sum_{n=1}^N{\tilde h}_n{\tilde {\bf p}}_n+\sum_{n=1}^{N-1}\sum_{l=1}^n({\tilde {\bf q}}_{n+1}-{\tilde {\bf q}}_l)\,{\tilde j}_{n+1,l}\,,
\end{equation}
with the dimensionless local energy densities and currents,
\begin{equation}
{\tilde h}_n=\frac{h_n}{\ell^2m\nu^2}
\end{equation}
and
\begin{equation}
{\tilde j}_{l,n}=\frac{j_{l,n}}{\ell^2m\nu^3}\,,
\end{equation}
respectively. The dimensionless local temperatures are given by
\begin{equation}
{\tilde T}_n=\left(\frac{K_B}{\ell^2m\nu^2}\right)T_n=\frac{1}{2}\sum_{\mu=\{x,y\}}<{\tilde p}_{\mu,n}^2>_{\varepsilon}\,.
\end{equation}

In the next section we will continue the analysis in terms of the dimensionless variables. We will remove the tilde symbol from all the variables and parameters to simplify the notation.

%%%%%%%%%%%%%%%%%%%%%%%%%%%%%%%%%%%%%%%%%%%%%%%%%

\section{The numerical integration}

The $4N-$dimensional stochastic differential equations (\ref{eqn3}) can be expressed in a compact form as
\begin{equation}\label{eq}
d{\bf Y}={\bf A}({\bf Y})\,dt\,+\,{\bf B}\cdot d\bm{\Omega}_t\,,
\end{equation}
where the components of the variable vector ${\bf Y}$ have been ordered in the form
\begin{equation}
{\bf Y}=(q_{x,1},\dots,q_{x,N},q_{y,1},\dots,q_{y,N},{p}_{x,1},\dots{p}_{x,N},{p}_{y,1},\dots,{p}_{y,N})\,.
\end{equation}
The components of the vector ${\bf A}$ containing the deterministic terms in the equations of motion are
\begin{eqnarray}
A_{i}=
\left\{
\begin{array}{ll}
{p}_{x,i}\,, & \quad  i=1,\dots,N\,,  \nonumber\\
{p}_{y,i-N}\,, & \quad i=N+1,\dots,2N\,,\nonumber\\
-\left(\frac{\partial {\cal V}}{\partial q_{x,i-2N}}+\eta_{x,i-2N}\,p_{x,i-2N}\right)\,, & \quad i=2N+1,\dots,3N\,,\nonumber\\
 -\left(\frac{\partial {\cal V}}{\partial q_{y,i-3N}}+\eta_{y,i-3N}\,p_{y,i-3N}\right)\,, & \quad i=3N+1,\dots,4N\,.
 \end{array}
 \right.
\end{eqnarray}
The matrix ${\bf B}$ contains the diffusion coefficients $D_{\mu,n}$. In our model it can be expressed by a diagonal matrix with the elements
\begin{eqnarray}
B_{ii}=
\left\{
\begin{array}{ll}
0\,, & \quad  i=1,\dots,N\,,  \\
0\,, & \quad i=N+1,\dots,2N\,,\\
\sqrt{2\,D_{x,i-2N}}\,, & \quad i=2N+1,\dots,3N\,,\\
\sqrt{2\,D_{y,i-3N}}\,, & \quad i=3N+1,\dots,4N\,.
 \end{array}
 \right.
\end{eqnarray}
The vector $d\bm{\Omega}_{t}$ denotes the $4N-$dimensional Wiener process, whose elements are
\begin{eqnarray}
d\Omega_{t,i}=
\left\{
\begin{array}{ll}
0\,, & \quad  i=1,\dots,2N\,,  \\
dW_{i-2N}\,, & \quad i=2N+1,\dots,3N\,,\\
dW_{i-3N}\,, & \quad i=3N+1,\dots,4N\,.
 \end{array}
 \right.
\end{eqnarray}
To integrate the stochastic differential equations (\ref{eq}) we consider the multi-dimensional explicit order 2.0 weak scheme proposed by Platen \cite{Kloeden}. Since in our model the matrix ${\bf B}$ does not depend explicitly on the variable ${\bf Y}$, such scheme  is particularly simple. Given the variable ${\bf Y}_{I}$ at a time step $I$, its value at the following time step $I+1$ is given by
\begin{equation}
{\bf Y}_{I+1}={\bf Y}_{I}+\frac{1}{2}\left[{\bf A}(\bm{\Gamma}_{I})+{\bf A}({\bf Y}_{I})\right]\Delta_{t}+{\bf B}\cdot\Delta\bm{\Omega}_{I}
\end{equation}
where
\begin{equation}
\bm{\Gamma}_{I}={\bf Y}_{I}+{\bf A}({\bf Y}_{I})\Delta_{t}+{\bf B}\cdot\Delta\bm{\Omega}_{I}
\end{equation}
$\Delta_{t}=t_{I+1}-t_{I}$ is the constant time interval between two consecutive time steps, and $\bm{\Omega}_{I}$ is the vector with elements
\begin{eqnarray}
\Delta\Omega_{I,i}=
\left\{
\begin{array}{ll}
0, & \quad i=1,\dots,2N\,,  \\
\sqrt{\Delta_{t}}\,\,G_{I,x,i-2N}, & \quad i=2N+1,\dots,3N\,,\\
\sqrt{\Delta_{t}}\,\,G_{I,y,i-3N}, & \quad i=3N+1,\dots,4N\,,
 \end{array}
 \right.
\end{eqnarray}
with $G_{I,\mu,n}\sim N(0;1)$ (standard Gaussian) a normally distributed random variable selected at time step $I$ for the $n$-ion, along the $\mu$-direction.

In order to get stable local temperatures and a total heat flux from the solutions of the equations (\ref{eq})  numerical simulations were performed considering time intervals $\Delta_t<1\times10^{-4}$, and integrated up to a final time in which the steady conditions (\ref{steady1}) and (\ref{steady3}) were satisfied. The process became computationally expensive as the multiple interaction potential (\ref{vad}) had to be evaluated more than $4\times10^7$ times for each stochastic, and averages that included over $500$ of these trajectories were considered. It took over two months to carry out the simulations and get stable results using a $32$ CPU  machine with AMD Opteron (tm) Processors 6134.

\section{Characteristic timescales in the non-equilibrium dynamics towards the steady-state}
%%%%%%%%%%%%%%%%%%%%%%%%%%%%%%%%%%%%%%%%%%%%%%%%%
\begin{figure}
\centering
\includegraphics[width=0.7\columnwidth]{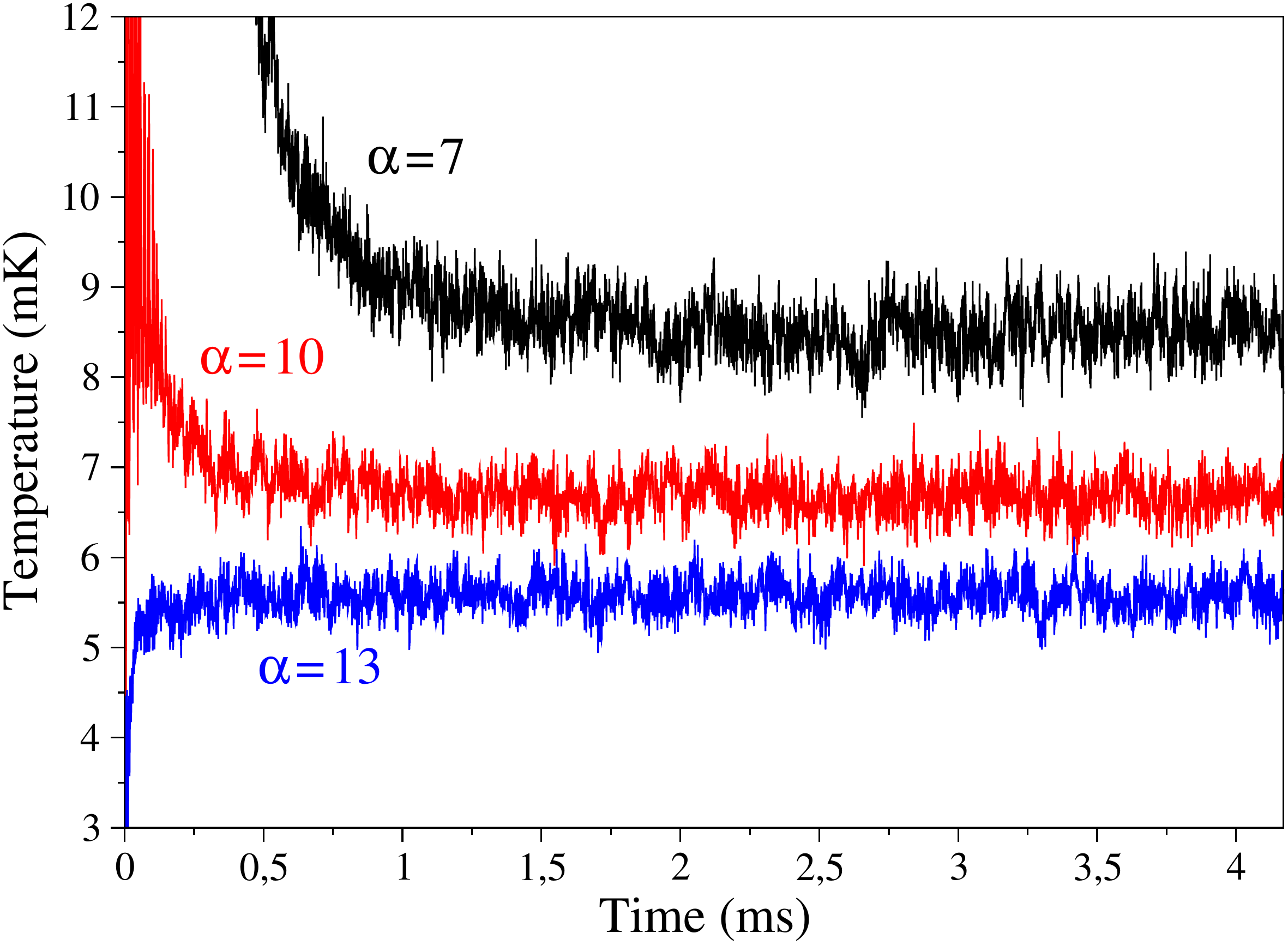}
\includegraphics[width=0.7\columnwidth]{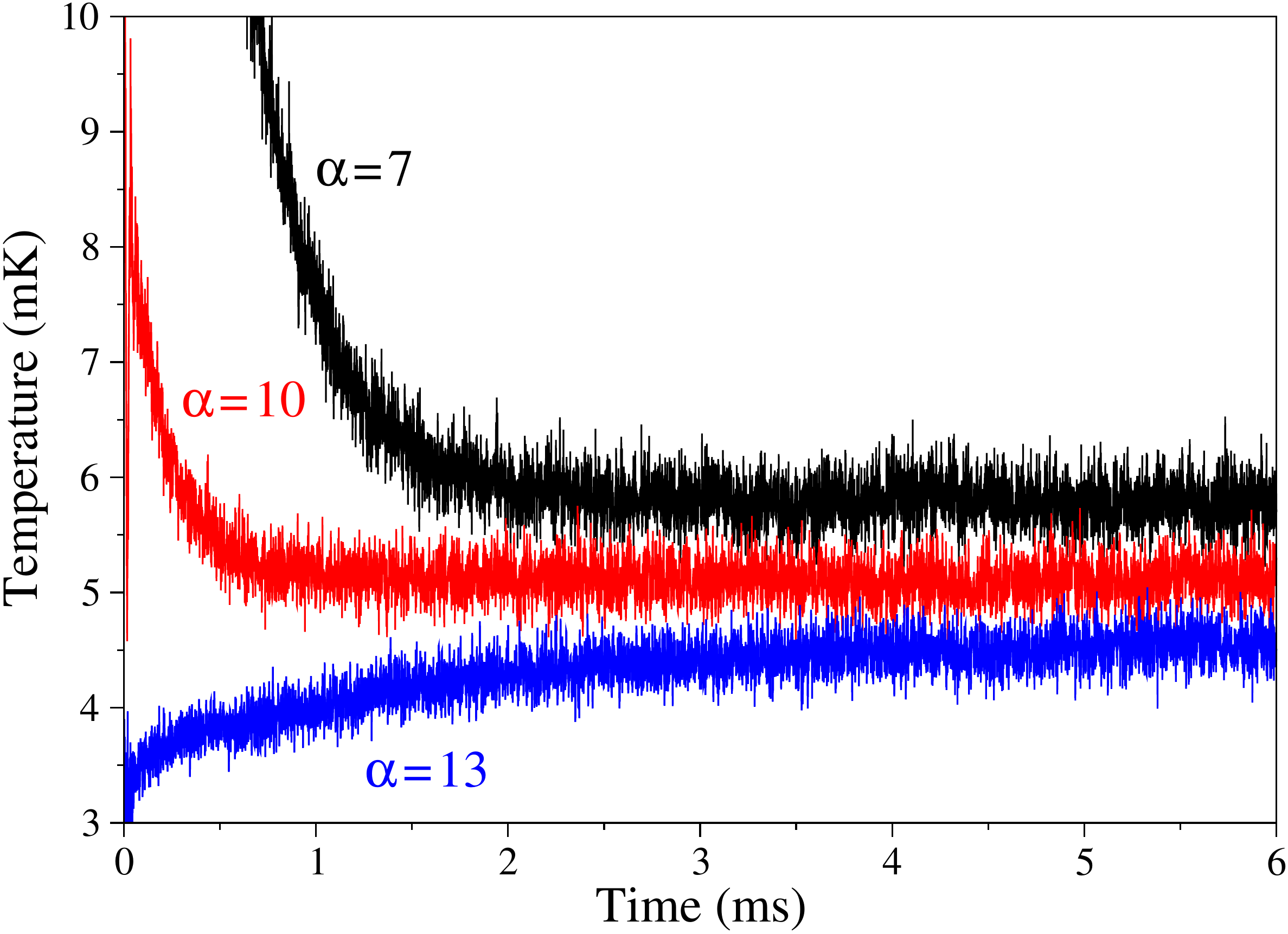}
\caption{
(Color online) Time evolution of the local temperatures of the leftmost ion ($n=1$, upper panel) and the central ion ($n=15$, lower panel) for chains with different ratios of the trap frequencies.
}
\label{fig_temp_time}
\end{figure}
%%%%%%%%%%%%%%%%%%%%%%%%%%%%%%%%%%%%%%%%%%%%%%%%%
So far we have presented results obtained in terms of dynamical variables collected from long enough simulations  for the system to reach a non-equilibrium steady state. In this section we indicate the characteristic timescales that are needed to achieve such state, with associated time-independent local temperatures and heat fluxes.
%%%%%%%%%%%%%%%%%%%%%%%%%%%%%%%%%%%%%%%%%%%%%%%%%
\begin{figure}[h]\centering
\includegraphics[width=0.75\linewidth]{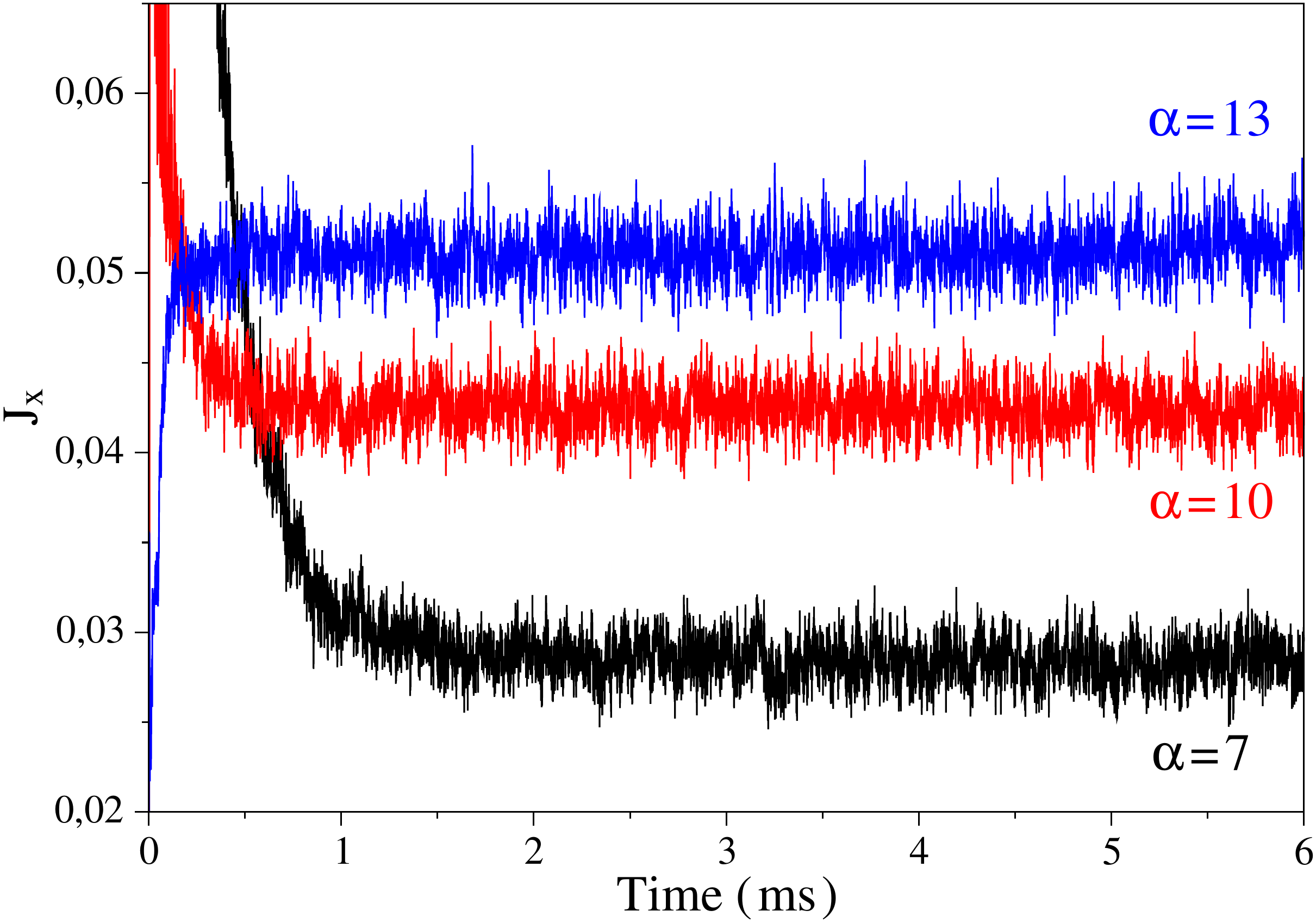}
\caption{
(Color online) Time evolution of the total heat flux in the axial direction for chains with different ratios of the trap frequencies.
}
\label{fig_Jb_time}
\end{figure}
%%%%%%%%%%%%%%%%%%%%%%%%%%%%%%%%%%%%%%%%%%%%%%%%%
Figures (\ref{fig_temp_time}) and (\ref{fig_Jb_time}) show some representative time evolutions of the local temperatures and the total heat fluxes towards the steady state configuration, obtained from an average over more than $800$ stochastic trajectories. The ions at both ends of the chain that are directly connected to the laser beams reach steady values faster than the inner ions. The slowest convergence occurs for the central ions of the linear chain. In general, it can be assumed that the system stabilizes after approximately $5-10\,$ms.
To ensure accurate temperature profiles and heat fluxes, the simulations have been performed up to $13\,$ms, and the final results have been obtained from a time average within the interval $[10,13]\,$ms.

Hence we are considering short-time-scale experiments in which effects such the motional heating of the trapped ions
confined in rf-traps due to fluctuating electric fields from the trap electrodes should not be relevant.

\end{document}